\documentclass[conference]{IEEEtran}
\usepackage{cite}
\usepackage{amsmath,amssymb,amsfonts}
\usepackage{algorithmic}
\usepackage{graphicx}
\usepackage{textcomp}
\usepackage{xcolor}
\def\BibTeX{{\rm B\kern-.05em{\sc i\kern-.025em b}\kern-.08em
    T\kern-.1667em\lower.7ex\hbox{E}\kern-.125emX}}
\usepackage{subfig}
\newsavebox{\measurebox}    
\newcommand{\norm}[1]{\left\lVert#1\right\rVert}
\newcommand{\R}{\mathbb{R}}
\usepackage{amsmath,amssymb,amsthm}
\usepackage{stmaryrd}

\usepackage{algorithm2e}
\usepackage{comment}
\begin{document}
%\title{Understanding the Spatio-temporal Topic Dynamics in Social Media using Non-negative Tensor Factorization}
\title{Understanding the Spatio-temporal Topic Dynamics of Covid-19 using Nonnegative Tensor Factorization: A Case Study}
%\author{\IEEEauthorblockN{1\textsuperscript{st} Thirunavukarasu Balasubramaniam}
%\IEEEauthorblockA{\textit{Centre for Data Science, \\School of Computer Science} \\
%\textit{Queensland University of Technology}\\
%Brisbane, Australia \\
%\and
%\and

%\author{Thirunavukarasu Balasubramaniam, Richi Nayak, Md Abul Bashar}
%\affiliation{%
%  \institution{Queensland University of Technology}
%  \city{Brisbane, Queensland, Australia}}
%\email{{m1.bashar, r.nayak, t.balasubramaniam}@qut.edu.au}

\author{\IEEEauthorblockN{Thirunavukarasu Balasubramaniam, Richi Nayak, Md Abul Bashar}
\IEEEauthorblockA{Centre for Data Science, School of Computer Science\\
Queensland University of Technology\\
Brisbane, Queensland, Australia\\
\{thirunavukarasu.balas, r.nayak, m1.bashar\}@qut.edu.au}
}

\maketitle
\begin{abstract}
Social media platforms facilitate mankind a data-driven world by enabling billions of people to share their thoughts and activities ubiquitously. This huge collection of data, if analysed properly, can provide useful insights into people's behavior. More than ever, now is a crucial time under the Covid-19 pandemic to understand people's online behaviors detailing what topics are being discussed, and where (space) and when (time) they are discussed. Given the high complexity and poor quality of the huge social media data, an effective spatio-temporal topic detection method is needed. This paper proposes a tensor-based representation of social media data and Non-negative Tensor Factorization (NTF) to identify the topics discussed in social media data along with the spatio-temporal topic dynamics. A case study on Covid-19 related tweets from the Australia Twittersphere is presented to identify and visualize spatio-temporal topic dynamics on Covid-19.

\end{abstract}

\begin{IEEEkeywords}
Covid-19, Social media analysis, Topic modelling, Nonnegative Tensor Factorization, Spatio-temporal pattern mining
\end{IEEEkeywords}

\section{Introduction}

%Para 1: What is Covid-19 and why social media mining is good
The Covid-19 pandemic in 2020 has brought a never seen challenge to humankind and has impacted our lives drastically. It has changed the way we live and interact with each other. With physical distancing and isolation, social media has become a common portal for communication. The recent statistics show that there has been a $30\%$ to $45\%$ increase in social media consumption during this period\footnote{https://blog.twitter.com/en\_us/topics/company/2020/An-update-on-our-continuity-strategy-during-Covid-19.html}. Understanding people’s behavior during this pandemic will become a basis for many social and economical studies.

%Para 2: What type of data? 
The data generated in social media platforms like Twitter are text-based contents (tweets) and they are usually associated with space (location) and time.  The topic modeling on the tweets can help identify emerging topics in the tweets, while space and time information can reveal the spatial and temporal dynamics of those topics~\cite{saha2012learning}. Identifying these saptio-temporal topic dynamics, that highlights people’s mindset, is beneficial to policymakers. It allows them to take necessary actions for the well-being of people.

%Para 3: What are the machine learning techniques? – How tensor factorization is suitable here?
Social media data includes unstructured phrases that result in a huge variance in traditional text vocabulary. Traditional topic modelling methods such as Latent Dirichlet Allocation (LDA)~\cite{blei2003latent} are challenged with short text data and face sparsity and low word co-occurrence issues. Factorization methods like Non-negative Matrix Factorization (NMF)~\cite{nmf} map the high-dimensional  (sparse) text  representation  to  a  lower-dimensional representation. These methods have become popular in text mining due to their capability to capture the patterns in the lower dimensional representation of the data~\cite{lee1999learning}. NMF decomposes a high-dimensional (tweet $\times$ term) matrix into two low-rank factor matrices that represent tweet and term clusters. It produces  a  part-based  representation  by allowing  only  additive  combinations  of basis components~\cite{lee1999learning}. 

Non-negative Tensor Factorization (NTF)~\cite{kolda2009tensor}, an extension of NMF for high-dimensional data, can identify associations among multiple dimensions~\cite{sun2016understanding}. This brings an added advantage to NTF over NMF as the patterns can be interpreted with associations. NTF based methods have been used for spatio-temporal patterns elicitation on traditional (i.e. structured) data. %However, to elicit the spatio-temporal patterns along with topics from the social media data, a proper data representation is essential. 
%Additional contexts and constraints need to be introduced to NTF to discover more accurate patterns~\cite{balapatternmining2019,balasubramaniam2020column}. 
NTF for spatio-temporal pattern elicitation on social media (i.e. unstructured) data brings multiple challenges. %Firstly, the representation of text data along with spatio-temporal information in tensor. 
Firstly, representing this unstructured text data and spatio-temporal information in a single tensor representation model is challenging when there is a need to preserve the association among them. Secondly, the short texts from social media like Twitter can induce sparseness to the tensor representation. The state-of-the-art factorization algorithms may fail to effectively learn the spatio-temporal patterns in the presence of noise and sparsity present in social media data~\cite{oktar2018review,balapatternmining2019}. 
Finally, the larger data size introduces efficiency issues in factorization process~\cite{balasubramaniam2020column}. Therefore, a proper data representation and selection of a suitable factorization algorithm is crucial to deal with social media data.

%Para 4: Contributions of this paper
In this paper, we present a novel NTF based spatio-temporal topic dynamics
discovery method. We focus on applying the best-suited data representation model and the factorization algorithm to understand the spatio-temporal distribution of topics emerging from users' interactions on social media related to Covid-19. 
%The key contributions in this paper are as follows. Firstly, we propose to represent social media data, especially the tweets, in a tensor representation along with the spatial and temporal contexts. %Secondly, the tensor represented data is factorized using an advanced factorization technique that can  and the factorization output (factor matrices) is used to elicit spatio-temporal topic dynamics from social media. 
%Secondly, the tensor represented data is factorized using a factorization technique that can handle the large volume of social media data and facilitate high-quality spatio-temporal pattern elicitation from sparse data. Finally, the elicited patterns are visualized for easy understanding of patterns and to identify the association between the spatio-temporal patterns and the topics. 
We present a case study analysing a large tweet collection of the Australian Twittersphere\footnote{Location of author or tweet or a location mentioned in the tweet is Australia or any of its cities} containing certain keywords relating to Covid-19. Extensive insights explaining the spatio-temporal topic dynamics around Covid-19 have been identified. To our best of knowledge, for the first time, this paper demonstrates the capability of NTF in discovering the spatio-temporal topic dynamics in social media.

\section{Related work}
The massive amounts of data are generated by users in social media platforms nowadays and mining it for insights can bring many advantages in many domains~\cite{gundecha2012mining}. For instance, authors in \cite{culotta2010towards} attempted to detect influenza outbreak by analyzing 500,000 Twitter messages. It is based on a predictive model that classifies the Twitter messages with flu-related terms like fever, cough, and sore throat to a potential influenza discussion. In a recent study \cite{han2020using} related to Covid-19, the social media data from Sina-Weibo which is Twitter-like short text was analyzed using the LDA, a topic modeling technique by which the common topics in the texts were identified. These kinds of insights are timely and can help the government and emergency authorities to act effectively, and are socially important. In addition to the topics, the identification of spatio-temporal patterns is also vital to get a broader understanding of the topics. %In \cite{resch2018combining}, the topics patterns related to earthquakes are identified using LDA and the spatio-temporal patterns are identified by classifying the tweets based on the topics. While the topic-spatial and topic-temporal association can be captured using this approach, it fails to capture the topic-spatial-temporal association.

%Factorization for spatio-temporal pattern mining
Spatial and temporal pattern mining is a much studied research area and there exist multiple factorization based techniques\cite{gao2019spatiotemporal,ma2018identifying,fan2014cityspectrum,balasubramaniam2019sparsity}. NMF has been extensively used to understand the spatial or temporal patterns independently. In \cite{gao2019spatiotemporal}, the location-time information of taxi trips is represented in a matrix and NMF is used to discover the spatio-temporal patterns. With a similar representation, authors in\cite{ma2018identifying} applied NMF to identify the traffic patterns from the road-network data. While all of these techniques are capable of extracting spatial or temporal patterns existing in the data, similar to LDA, they can not associate the spatial and temporal patterns concurrently with the additional dimension like users or drivers. This is due to the matrix representation capability which supports a two-dimensional association only.

%Nonnegative Tensor Factorization for spatio-temporal pattern mining
NTF based methods can represent the associations between spatial and temporal patterns due to the NTF's ability to represent the multi-dimensional data with multi-dimensional associations preserved. In~\cite{fan2014cityspectrum}, the Tokyo city mobility data is represented as a tensor form (region $\times$ day $\times$ time) and NTF is used to understand the mobility patterns involving spatial and temporal behaviors. In~\cite{takeuchi2017structurally}, a structurally regularized NTF is proposed for spatio-temporal pattern discoveries in the traffic flow data. In another work, a modified NTF with sparsity constraint is proposed for the extraction of spatio-temporal patterns with reduced noise~\cite{balasubramaniam2019sparsity}. Recently, tensor modelling with NTF was proposed to extract spatio-temporal patterns of Singapore's elderly people using the multi-context data collected using sensors~\cite{balasubramaniam2020column}.  

While NTF has been used in many applications generating spatio-temporal patterns in the last 5 years, it has rarely been used to generate topics based on spatio-temporal patterns from social media data. This is mainly because of three challenges. 1) Representation of text data along with spatio-temporal information in the tensor model. 2) The NTF on sparse data tends to produce patterns with a simultaneous elimination problem (inability to avoid learning repetitive patterns)~\cite{balasubramaniam2019sparsity}. 3) The social media data is a huge collection and NTF is computationally expensive due to multiple complex matrices and tensor products involved. 

As the spatio-temporal topic dynamics of social media is itself a fairly new research area~\cite{resch2018combining}, and the development of NTF for spatio-temporal pattern mining is also emerging, this paper is a first attempt to bridge this gap. This paper proposes a novel method based on a tensor representation of social media data and solving it using NTF for understanding spatio-temporal topic dynamics.

\section{NTF based spatio-temporal topic discovery}
The proposed NTF based spatio-temporal topic dynamics discovery method consists of the following three components: 1) Data representation, 2) Non-negative Tensor Factorization, and 3) Spatio-temporal topic dynamics.%, as shown in Fig.~\ref{arch}.
\begin{comment}
\begin{figure*}[htb!]
\centering
\includegraphics [width=5.2in, height = 2.8in]{images/arch.pdf}
\caption{Architecture: NTF based spatio-temporal topic dynamics discovery}
\label{arch}
\end{figure*}
\end{comment}
\subsection{Data representation}
The social media exhibit the data in multifaceted nature~\cite{darmon2015followers}. Each tweet (collection of terms) is associated with additional contexts like time of the tweet and location of the user. Therefore, a total of three attributes, terms, time, and location has to be simultaneously analyzed for identifying common patterns. 

\subsection*{Existing Data Models: Matrix representation}
\begin{figure}[]
    \centering
    \subfloat[term $\times$ location matrix]{{
     \includegraphics[width=2.3cm, height=2.6cm]{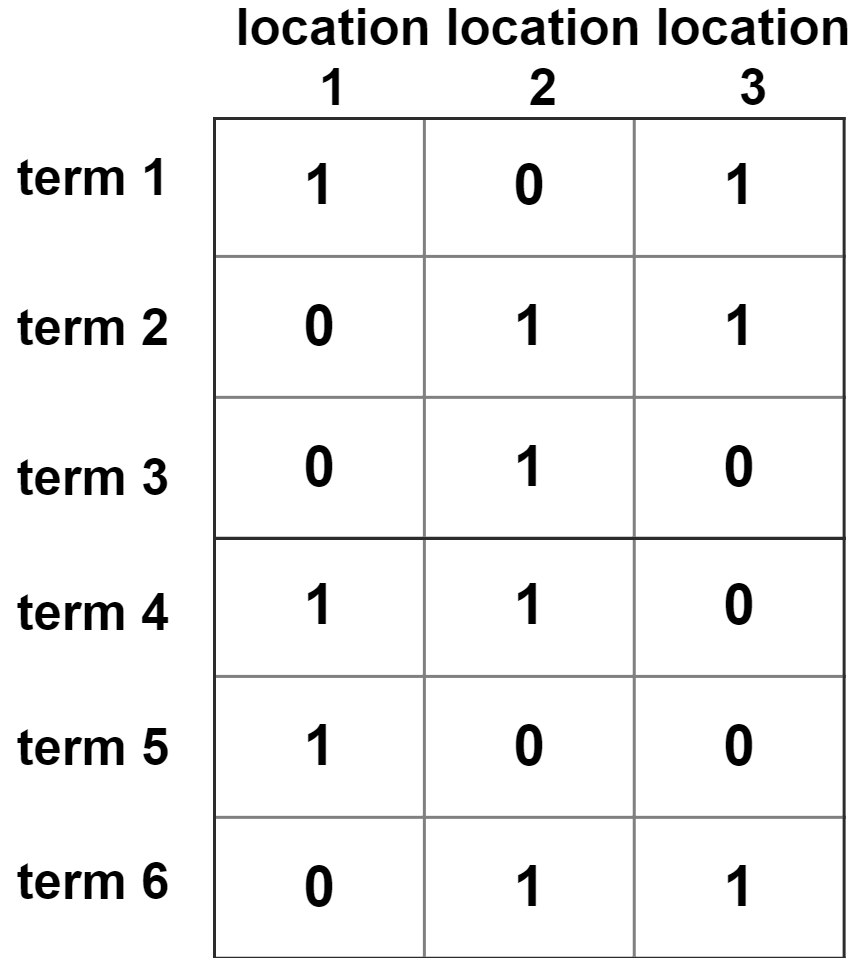} 
    }}%
    \qquad
    \subfloat[term $\times$ time matrix]{{
     \includegraphics[width=2.3cm, height=2.6cm]{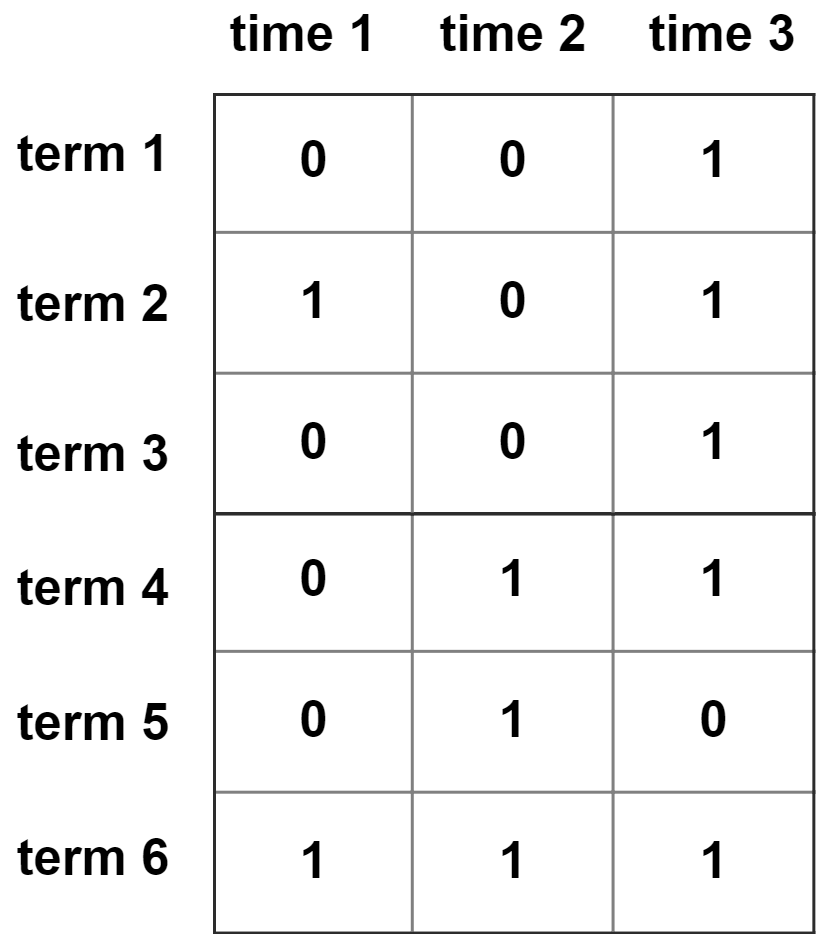} 
    }}%
    \caption{(a) term $\times$ location, and (b) term $\times$ time matrix representation of tweets; 0 indicates the absence of term and  non-zero entries indicate the frequency of the term.}%
    \label{fig:matrixrep}%
\end{figure}

The existing methods represent this information in a matrix form, e.g., (term $\times$ location) and (term $\times$ time) matrices. Fig.~\ref{fig:matrixrep} shows a toy example with 6 unique terms representing a set of tweets, associated with 3 unique locations and 3 unique time slots. The time can be set depending on the type of analysis conducted, such as day, week, and year.     

From the matrix representation, as shown in Fig.~\ref{fig:matrixrep}, the spatial and temporal patterns can be learned using NMF. While this kind of pattern mining is the state-of-the-art~\cite{gao2019spatiotemporal,ma2018identifying,fan2014cityspectrum,balasubramaniam2019sparsity}, it does have a disadvantage of losing association among location and time dimensions. For example, the spatial patterns learned from Fig.~\ref{fig:matrixrep}a can help to devise the collection of common terms used in specific locations, however, it fails to learn the associations such as how the collection of terms are used in different locations for a different time. This is due to the independent handling of location and time under two different matrix representations. A high-dimensional representation of this data is essential to identify patterns with the capability of learning the associations among all the modes or dimensions (term, location, and time).

In pandemic situations like Covid-19, it is useful to learn the dependencies of the spatial and temporal patterns. %For instance, a specific temporal pattern can be noticed in a specific country (i.e., location). But it may not be representing all the locations.
For instance, a specific temporal pattern may be noticed only in a specific country (i.e. location) and not necessarily noticed in all the countries as a common pattern.  Therefore, learning the association within the time and location is important. Consider a real example. During the early days of the Covid-19 outbreak, conversations revolve around China, a single location. However, this has changed over time. Currently, it is widespread and each country is having a different time period of Covid-19 peak. Therefore, patterns highlighting the proper association of time and location will only be useful.  %Due to the extreme importance of learning the associations among all the modes, the tensor model is proposed 

\subsection*{Proposed Tensor based data model}
To address this, we propose to use the tensor model to represent multiple dimensions inherent with social media data. Tensor is an extension of vector to present higher-order data that describes the elements of the multi-linear space. We present a three-dimensional tensor model to record the term, location, and time in a single representation by preserving the association among them.

%Vector is a tensor of order 1 and the matrix is a tensor of order 2. The structure greater than 2 is called a higher-order tensor.
Let $\mathcal{U}$ = \{$u_1,u_2,...,u_M$\}, $\mathcal{L}$ =  \{$l_1,l_2,...,l_N$\} and $\mathcal{T}$ =  \{$t_1,t_2,...,t_O$\} be the set of unique terms, locations and time periods in the tweets collection, respectively. Consider term 4 from Fig.~\ref{fig:matrixrep} that is associated with locations 1, 2 and times 2, 3. In the matrix representation, it is unclear that at what time, term 4 associated with location 1. Suppose, term 4 is associated with locations 1 and 2 at time period 2, this can be captured using a tensor model as, $\boldsymbol{\mathcal{X}}_{4,1,2}$ = $(u_4,l_1,t_2)$ and $\boldsymbol{\mathcal{X}}_{4,2,2}$ = $(u_4,l_2,t_2)$, where $\boldsymbol{\mathcal{X}}$ is the tensor of size  $\R^{(M \times N \times O)}$. Instead of representing the data with two independent matrices as in Fig.~\ref{fig:matrixrep},  we represent the data with a single multidimensional tensor representation. 

\subsection{Non-negative Tensor Factorization}
Once the data is represented as a tensor model, the spatio-temporal topic dynamics can be learned using NTF. Factorization, a dimensionality reduction technique, is the process of decomposing the high-dimensional data into factor matrices learning the dependencies within the data. %The factor matrices (i.e., the lower-dimensional representation) are a building block for many applications like pattern mining~\cite{maruhashi2011multiaspectforensics,naveh2018urban}, text mining~\cite{chen2018adaptive,li2013nonnegative}, and recommender systems~\cite{haque2019divergence,ioannidis2019coupled}.
Candecomp/Parafac (CP)~\cite{carroll1970} factorization decomposes the tensor into multiple rank-1 tensors.% as shown in Fig.~\ref{lrfig:cpf}. %For a third-order tensor $\boldsymbol{\mathcal{X}} \in \R^{(M \times N \times O)}$, each rank-1 tensor comprises of three components representing each mode of the tensor. In CP, there is no core tensor available and therefore, it is less complex when compared to Tucker~\cite{kolda2009tensor}. The mode-wise components are grouped to form the factor matrices representing each mode of the tensor.

The CP factorization can be defined as,
\begin{equation}
\label{lreq:cp factorization}
   \boldsymbol{\mathcal{X}} \cong \llbracket\boldsymbol{\mathrm{U}}, \boldsymbol{\mathrm{L}}, \boldsymbol{\mathrm{T}} \rrbracket = \sum_{r=1}^{R} \boldsymbol{\mathrm{u_{r}}} \circ \boldsymbol{\mathrm{l_{r}}} \circ \boldsymbol{\mathrm{t_{r}}},
\end{equation}
where \(\boldsymbol{\mathrm{U}} \in \R^{(M \times R)}\), \(\boldsymbol{\mathrm{L}} \in \R^{(N \times R)}\) and \(\boldsymbol{\mathrm{T}} \in \R^{(O \times R)}\) are factor matrices with \(R\) hidden features (rank), \(R \in \mathbb{Z}_{+} \). $\boldsymbol{\mathrm{u_{r}}}$, $\boldsymbol{\mathrm{l_{r}}}$ and $\boldsymbol{\mathrm{t_{r}}}$ are the $r^{th}$ column of $\boldsymbol{\mathrm{U}}$, $\boldsymbol{\mathrm{L}}$ and $\boldsymbol{\mathrm{T}}$ respectively.  Unlike Tucker factorization, the CP factorization should have the same number of rank value for all the factor matrices.

The optimization minimization problem of CP factorization can be defined as,
\begin{equation}
    \label{lreq:cpopt}
    f(\boldsymbol{\mathrm{U}},\boldsymbol{\mathrm{L}},\boldsymbol{\mathrm{T}}) = \norm{\boldsymbol{\mathcal{X}} - \llbracket\boldsymbol{\mathrm{U}}, \boldsymbol{\mathrm{L}}, \boldsymbol{\mathrm{T}} \rrbracket}^2,
\end{equation}
where \(\boldsymbol{\mathrm{U}} \in \R^{(M \times R)}\), \(\boldsymbol{\mathrm{L}} \in \R^{(N \times R)}\) and \(\boldsymbol{\mathrm{T}} \in \R^{(O \times R)}\) are factor matrices with \(R\) hidden features (rank), \(R \in \mathbb{Z}_{+} \). $\boldsymbol{\mathrm{u_{r}}}$, $\boldsymbol{\mathrm{l_{r}}}$ and $\boldsymbol{\mathrm{t_{r}}}$ are the $r^{th}$ column of $\boldsymbol{\mathrm{U}}$, $\boldsymbol{\mathrm{L}}$ and $\boldsymbol{\mathrm{T}}$ respectively.
%This optimization minimization problem can be solved using any optimization algorithm~\cite{lee2001algorithms}. 

%Tensor factorization based on CP factorization can generate factor matrices that have negative values in it. Those negative values are avoided and left unused in applications such as pattern mining. These discarded values can be helpful if learned properly to generate positive values instead of negative values~\cite{lee1999learning,welling2001positive}. To make the tensor factorization into an NTF, a non-negative constraint is imposed on Eq.~\eqref{lreq:cpopt}. The basic idea of non-negative constraint factorization is factorizing the matrix or tensor into factor matrices that follow the non-negative constraints~\cite{friedlander2008computing}. %non-negative constraint factorization can be achieved by traditional factorization methods by imposing constraints to maintain the non-negative values in the output. 
It has been shown that introducing non-negative constraints on factor matrices will improve the accuracy of the approximation of matrix and tensor~\cite{nmf}. The optimization minimization function of CP-based NTF can be defined based on~\eqref{lreq:cpopt} as, 
\begin{equation}
    \label{lreq:ntf}
    f(\boldsymbol{\mathrm{U}},\boldsymbol{\mathrm{L}},\boldsymbol{\mathrm{T}}) = \norm{\boldsymbol{\mathcal{X}} - \llbracket\boldsymbol{\mathrm{U}}, \boldsymbol{\mathrm{L}}, \boldsymbol{\mathrm{T}} \rrbracket}^2,~s.t.,~\boldsymbol{\mathrm{U}}\geq 0,~\boldsymbol{\mathrm{L}}\geq 0,~\boldsymbol{\mathrm{T}}\geq 0.
\end{equation}

The next task is to solve the given optimization minimization problem using a factorization algorithm. Since Eq.~\eqref{lreq:ntf} is a non-convex optimization, the Alternating Least Squares (ALS) algorithm which solves the one-factor matrix by fixing the other factor matrix is ideal~\cite{cichocki2009nonnegative}. However, due to its poor convergence speed and high memory requirement, Coordinate Descent (CD) based algorithms are proposed as alternatives~\cite{cicho2009fhals}. Moreover, the usage of existing factorization algorithms like ALS is not ideal for the sparse and multi-dimensional social media data as they are prone to generate denser factor matrices that include noise. This can negatively affect the quality of patterns learned from the sparse input.

%Next, we introduce a recent CD-based factorization algorithm, Saturating Coordinate Descent (SaCD)~\cite{balasubramaniam2020efficient} to deal with this data. 
Unlike other factorization algorithms, a recent CD-based factorization algorithm, Saturating Coordinate Descent (SaCD)~\cite{balasubramaniam2020efficient} is capable of learning the factor matrices with minimal noise, and hence the quality of patterns is relatively higher. SaCD is formulated in such a way that it can solve the NTF optimization problem efficiently for the social media data that is large in size and sparser in representation. %We use SaCD to learn the latent association among the three dimensions of term, time and location.

\subsection{Spatio-temporal topic dynamics using NTF}
\begin{figure}
\centering
\includegraphics [width=3.5in, height = 2.4in]{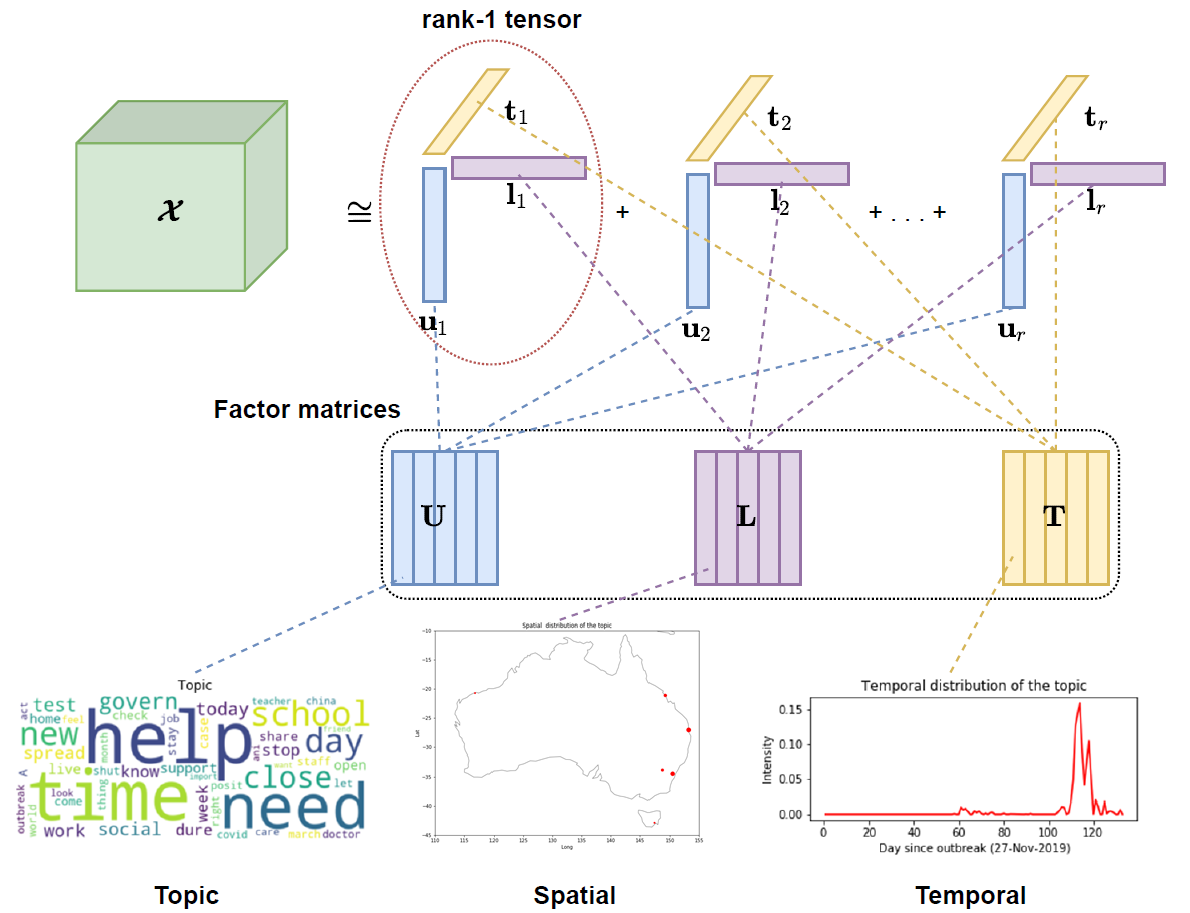}
\caption{Spatio-temporal patterns of topics using NTF output. Each column in a factor matrix represents a pattern.}
\label{cp_all}
\end{figure}

The factorization process using SaCD factorizes the input tensor $\boldsymbol{\mathcal{X}}$ into factor matrices $\boldsymbol{\mathrm{U}}$, $\boldsymbol{\mathrm{L}}$, and $\boldsymbol{\mathrm{T}}$. Each factor matrix represents the lower-dimensional representation of a mode (i.e., term/location/time) of $\boldsymbol{\mathcal{X}}$. Each column in a factor matrix represents a feature/component as shown in Fig.~\ref{cp_all}. The features are learned such that they reveal the latent relationships in the data and the values present in the feature show the strength of each mode's membership towards that feature~\cite{lee1999learning}. The rank value ($R$) assigned during factorization determines the number of features and each feature learns different relationships. For example, the factor matrix representing the time mode ($\boldsymbol{\mathrm{T}}$) consists of columns/features that represent the relationships of each hour concerning the terms and locations. For each hour, a distinct relationship is possible. Hence, each of them can be considered a temporal pattern. 

%The (term $\times$ location $\times$ time) tensor is populated with the observations that represent the number of times the term appeared in a specific location and time. After factorization of this tensor, three factor matrices ($\boldsymbol{\mathrm{U}}$, $\boldsymbol{\mathrm{L}}$, and $\boldsymbol{\mathrm{T}}$) are learned, with each factor representing a mode (or dimension). Each column of a factor matrix reveals a distinct pattern for the respective mode. 

While the features of location and time modes represent the spatial and temporal patterns respectively, the features of the term mode represent the topics. The topics are differentiated based on the values assigned to the terms. In each topic, a different set of terms will appear with high weights, which states that those terms are representative of the topic. For the example given in Fig.~\ref{cp_all}, the representative terms are help, need, govern, school, time, day, close. This gives insights on a topic. For example, this topic in Fig.~\ref{cp_all} is about the closer of schools and the need for government support. Similarly, the spatial and temporal patterns are differentiated based on the weight each location and time is assigned for each column. In the same example, the selected column from the temporal factor matrix $\boldsymbol{\mathrm{T}}$ has the high weights towards the end. And, the Australian cities like Sydney and Brisbane have a high weight as shown in the same column of location factor matrix $\boldsymbol{\mathrm{L}}$. It shows that the popularity of the topic is high in these cities at the given temporal pattern.

An important aspect that brings the advantage to NTF over NMF is the formation of rank-1 tensors as shown in Fig.~\ref{cp_all}. Each rank-1 tensor is a collection of the same rank features of all the modes; and the linear combination of these rank-1 tensors for all the rank is the approximation of the input tensor. This formation of rank-1 tensors makes the factorization algorithm to learn the features to be associated with each other. For example, the rank-1 tensor of $r^{th}$ feature/rank will associate the $r^{th}$ feature of all modes and hence they collectively represent the $r^{th}$ feature of the input. Therefore, the spatio-temporal patterns from location and time mode can be associated with topics from term mode.

\section{Case Study}
\subsection{Dataset}
The dataset used in this case study on Covid-19 consists of $2.7~million$ tweets (tweets that contains at least one of the following keywords: coronavirus, covid19, covid-19, covid\_19, coronovirusoutbreak, covid2019, covid, and coronaoutbreak) from Australian Twitter users that is collected from the start of Covid-19 (November $27^{th}$, 2019) to April $7^{th}$, spanning a timeline of $133~days$. Tweets are associated with multiple attributes and to understand spatio-temporal dynamics for each tweet, the respective user location and time are collected. The tweets with non-English languages and stop-words are removed for a better understanding of the output. The location of the user consists of many non-location entities and the proper location name is extracted using Named Entity Recognition (NER)~\cite{finkel2005incorporating}. For the tweets that do not have user location information, the tweets are searched for any location entity within them to populate missing locations. 

Using the unique values, a tensor of size ($15637~terms \times 1568~locations \times 133~days$) is populated with the count of a term's appearance at a given location at a given time. A rank $R = 10$ has been set to identify 10 spatio-temporal patterns of 10 topics. Setting a $R$ value higher $>$10 gives repetitive patterns and lower $<$10 misses the distinctive patterns. 

\subsection{Spatio-temporal patterns: NMF vs NTF}
\begin{figure*}[]
	%\centering
		\vskip -2.0\baselineskip plus -1fil
	\subfloat[Temporal pattern.]{{\includegraphics[width=2.3in, height=1.7in]{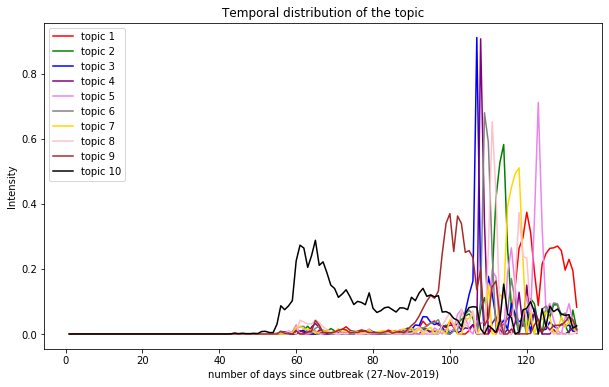}}}%
	%\qquad
	\subfloat[Spatial pattern]{{\includegraphics[width=2.3in, height=1.7in]{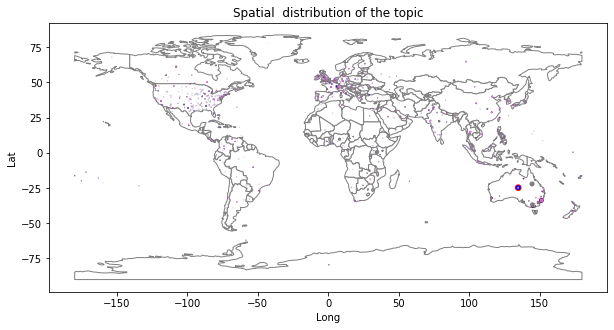}}}%
	%\qquad
	\subfloat[Spatial pattern within Australia]{{\includegraphics[width=2.3in,height=1.7in]{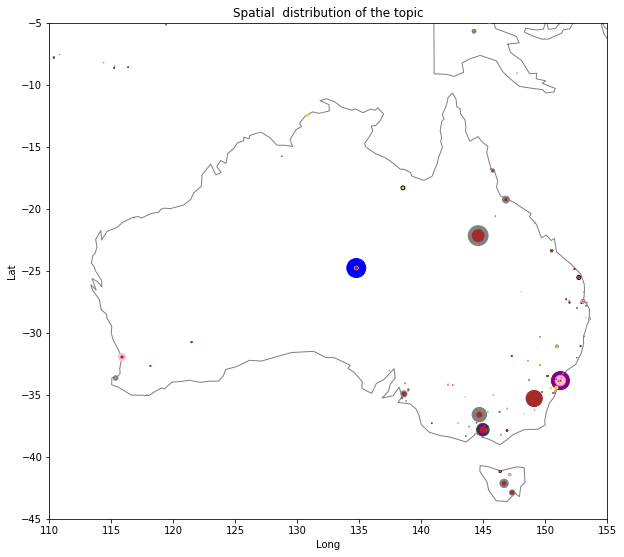}}}\hspace{5mm}%
	\caption{The spatio-temporal topic dynamics using NMF.}%
	\label{patterns_nmf}
\end{figure*}

Suppose NMF was used to extract the spatio-temporal patterns of topics identified from this dataset. Two matrices ($\boldsymbol{\mathrm{M1}}$ and $\boldsymbol{\mathrm{M2}}$) need to be created, one representing (term $\times$ time) and another representing (term $\times$ location) information. Applying NMF, temporal patterns of topics are learned from $\boldsymbol{\mathrm{M1}}$ (Fig.~\ref{patterns_nmf}a) and spatial patterns of topics are learned from $\boldsymbol{\mathrm{M2}}$ (Fig.~\ref{patterns_nmf}b). The spatial patterns within Australia are presented in Fig.~\ref{patterns_nmf}c. 

On the other hand, the same data can be represented as a single tensor (term $\times$ time $\times$ location) and the spatio-temporal patterns can be learned simultaneously from it. Fig.~\ref{patterns_ntf} shows the spatio-temporal patterns of topics learned using NTF. 

The advantages of NTF over NMF can be seen by trying to interpret these two figures. For instance, the black topic (Topic 10) seen in Fig.~\ref{patterns_nmf}a is the earliest temporal pattern. However, it cannot be interpreted from Fig.~\ref{patterns_nmf}b which are the locations associated with this pattern. This is because the spatial patterns are not learned in association with temporal patterns. Instead, they are learned independently along with the terms represented in two different matrices. 

In summary, NMF on $\boldsymbol{\mathrm{M1}}$ and $\boldsymbol{\mathrm{M2}}$ can learn the topics and its associated temporal and spatial patterns respectively, but not the both. Whereas, the spatio-temporal patterns learned from NTF can be associated. For example, the yellow topic (Topic 7) in Fig.~\ref{patterns_ntf}a is the earliest temporal pattern identified and they can be associated with the yellow spots in Figs.~\ref{patterns_ntf}b and \ref{patterns_ntf}c. Next, we will indulge in discussing the topics and patterns learned using NTF in detail.

\subsection{Patterns identified by NTF}
\begin{figure*}[]
	%\centering
		\vskip -2.0\baselineskip plus -1fil
	\subfloat[Temporal pattern.]{{\includegraphics[width=2.3in, height=1.7in]{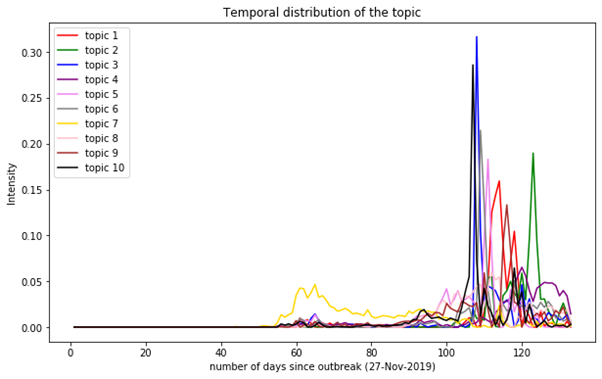}}}%
	%\qquad
	\subfloat[Spatial pattern]{{\includegraphics[width=2.3in, height=1.7in]{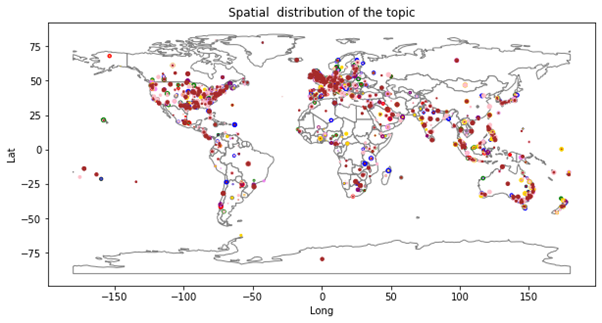}}}%
	%\qquad
	\subfloat[Spatial pattern within Australia]{{\includegraphics[width=2.3in,height=1.7in]{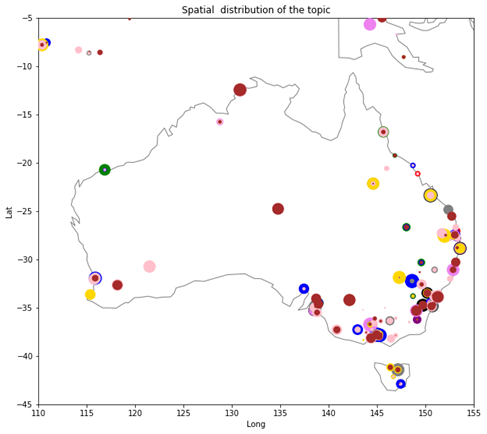}}}\hspace{5mm}%
	\caption{The spatio-temporal topic dynamics using NTF that can associate between the spatial and temporal patterns.}%
	\label{patterns_ntf}
\end{figure*}

Fig.~\ref{patterns_ntf} highlights the following insights:

\begin{enumerate}
    \item People in Australia did not tweet much for the first 60 days (until Jan 4$^{th}$ week) from the first known case of Covid-19.  
    \item Topic 7 (yellow color) is the first pattern that has been found in the early days. All other patterns approximately get started only from the 90$^{th}$ day (Feb 4$^{th}$ week).
    \item The active period is from the 100$^{th}$ day to 120$^{th}$ day (the month of March) and then it gets less intense.
    \item Topic 3 has seen only one high peak approximately on the 105$^{th}$ day (March 2$^{nd}$ week). 
\end{enumerate}

All the topics and their spatial patterns are shown in Fig.~\ref{patterns_ntf}b. Some interesting observations are: 1) Topic 9 (brown) is more spread across the world (Australian Twitter users tweeting from these locations) when compared to other topics. 2) Some countries in Africa shows only the Topic 7 (yellow) pattern.

\begin{comment}
\begin{figure}[!t]
\centering
\includegraphics [width=3in, height = 2.3in]{images/temp_all.png}
\caption{Temporal Pattern.}
\label{temp_all}
\end{figure}
\begin{figure}[!t]
\centering
\includegraphics [width=3in, height = 2.3in]{images/spatio_all.png}
\caption{Spatial Pattern of Australian Twitter users tweeting from worldwide locations.}
\label{spatio_all}
\end{figure}
\begin{figure}[!t]
\centering
\includegraphics [width=3in, height = 2.3in]{images/spatio_all_aus.png}
\caption{Spatial Pattern in Australia.}
\label{spatio_all_aus}
\end{figure}
\end{comment}

\begin{figure*}[ht!]
\centering
\sbox{\measurebox}{%
  \begin{minipage}[b]{.33\textwidth}
  \subfloat
    []
    {\label{fig:topic1A}\includegraphics[width=.77\textwidth,height=3.4cm]{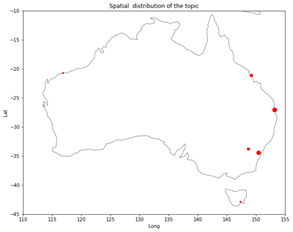}}
  \end{minipage}}
\usebox{\measurebox}\qquad
\begin{minipage}[b][\ht\measurebox][s]{.33\textwidth}
\centering
\subfloat
  []
  {\label{fig:topic1B}\includegraphics[width=.77\textwidth,height=1.5cm]{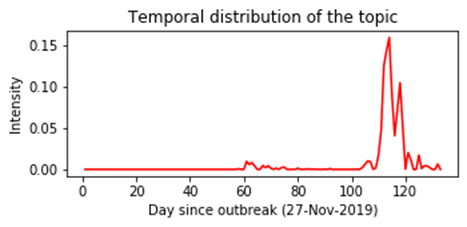}}
\vspace*{-1pt}
\subfloat
  []
  {\label{fig:topic1C}\includegraphics[width=.77\textwidth,height=1.5cm]{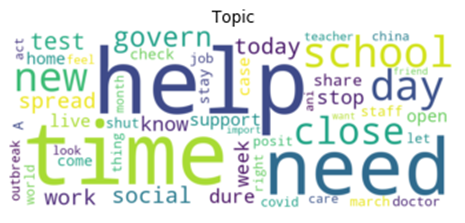}}
\end{minipage}
\caption{Topic 1. Summary: The closer of schools and the need for government support. Popular after 100 days and lasts for 20 days. Mainly centered in Sydney and Brisbane.}
\label{fig:topic1}
\end{figure*}

\begin{figure*}[ht!]
\centering
\sbox{\measurebox}{%
  \begin{minipage}[b]{.33\textwidth}
  \subfloat
    []
    {\label{fig:topic2A}\includegraphics[width=.77\textwidth,height=3.4cm]{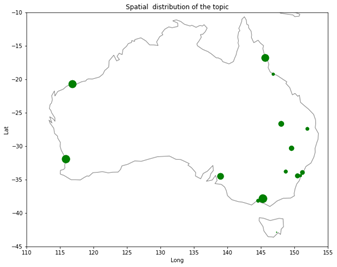}}
  \end{minipage}}
\usebox{\measurebox}\qquad
\begin{minipage}[b][\ht\measurebox][s]{.33\textwidth}
\centering
\subfloat
  []
  {\label{fig:topic2B}\includegraphics[width=.77\textwidth,height=1.5cm]{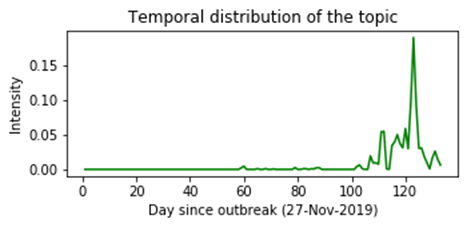}}
\vspace*{-1pt}
\subfloat
  []
  {\label{fig:topic2C}\includegraphics[width=.77\textwidth,height=1.5cm]{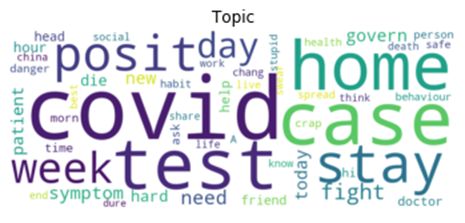}}
\end{minipage}
\caption{Topic 2. Summary: The start of more Covid-19 positive test results and stay at home slogan emerges. Popular after 120$^{th}$ day to 125$^{th}$ day. Mainly centered in Melbourne, Perth, Cairns, and Adelaide. While it is normal to expect Sydney in the list, it is not captured here as it has a different temporal pattern and more topics to it. Refer Topic 5 in Fig.~\ref{fig:topic5}.}
\label{fig:topic2}
\end{figure*}

\begin{figure*}[ht!]
\centering
\sbox{\measurebox}{%
  \begin{minipage}[b]{.33\textwidth}
  \subfloat
    []
    {\label{fig:topic3A}\includegraphics[width=.77\textwidth,height=3.4cm]{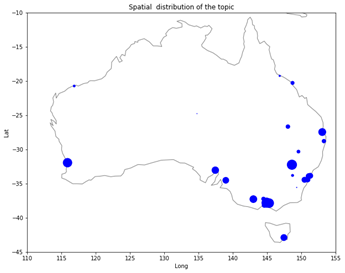}}
  \end{minipage}}
\usebox{\measurebox}\qquad
\begin{minipage}[b][\ht\measurebox][s]{.33\textwidth}
\centering
\subfloat
  []
  {\label{fig:topic3B}\includegraphics[width=.77\textwidth,height=1.5cm]{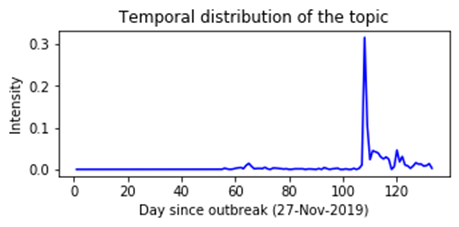}}
\vspace*{-1pt}
\subfloat
  []
  {\label{fig:topic3C}\includegraphics[width=.77\textwidth,height=1.5cm]{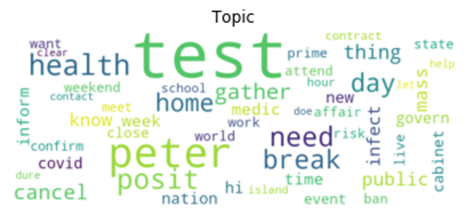}}
\end{minipage}
\caption{Topic 3. Summary: Peter Dutton, the Home Affairs minister of Australia tested positive for Covid-19. It is a short pattern that lasts only for a few days with a high peak for just one day. The topic is widely spread across all major cities in Australia.}
\label{fig:topic3}
\end{figure*}

\begin{figure*}[ht!]
\centering
\sbox{\measurebox}{%
  \begin{minipage}[b]{.33\textwidth}
  \subfloat
    []
    {\label{fig:topic4A}\includegraphics[width=.77\textwidth,height=3.4cm]{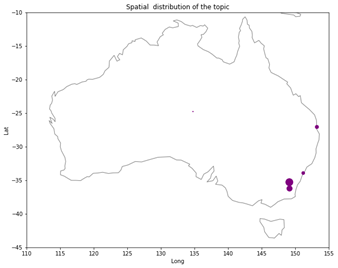}}
  \end{minipage}}
\usebox{\measurebox}\qquad
\begin{minipage}[b][\ht\measurebox][s]{.33\textwidth}
\centering
\subfloat
  []
  {\label{fig:topic4B}\includegraphics[width=.77\textwidth,height=1.5cm]{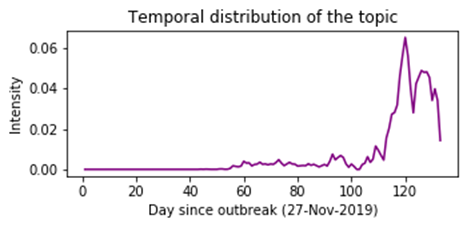}}
\vspace*{-1pt}
\subfloat
  []
  {\label{fig:topic4C}\includegraphics[width=.77\textwidth,height=1.5cm]{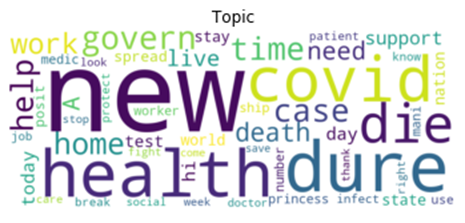}}
\end{minipage}
\caption{Topic 4. Summary: Reports on new cases and deaths associated with Covid-19. Most of the tweets are originated from the capital city of Canberra. This makes sense as the government and news media releases reports on Covid-19 cases from the capital territory.}
\label{fig:topic4}
\end{figure*}

\begin{figure*}[ht!]
\centering
\sbox{\measurebox}{%
  \begin{minipage}[b]{.33\textwidth}
  \subfloat
    []
    {\label{fig:topic5A}\includegraphics[width=.77\textwidth,height=3.4cm]{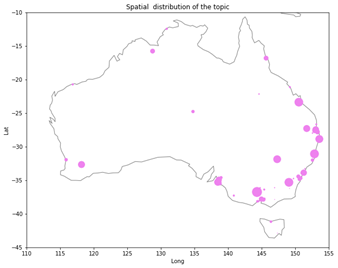}}
  \end{minipage}}
\usebox{\measurebox}\qquad
\begin{minipage}[b][\ht\measurebox][s]{.33\textwidth}
\centering
\subfloat
  []
  {\label{fig:topic5B}\includegraphics[width=.77\textwidth,height=1.5cm]{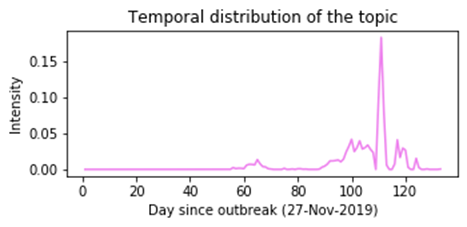}}
\vspace*{-1pt}
\subfloat
  []
  {\label{fig:topic5C}\includegraphics[width=.77\textwidth,height=1.5cm]{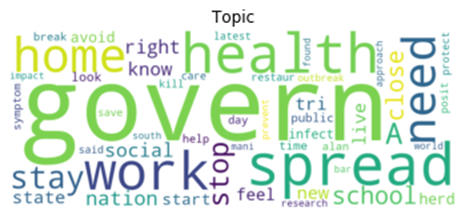}}
\end{minipage}
\caption{Topic 5. Summary: This topic can be considered as same as Topic 2, but the temporal pattern is approximately 10 days earlier than Topic 3. Moreover, the topic is spread across all major cities during this time. This also indicates how the same topic is temporally traveling across different locations.}
\label{fig:topic5}
\end{figure*}

\begin{figure*}[ht!]
\centering
\sbox{\measurebox}{%
  \begin{minipage}[b]{.33\textwidth}
  \subfloat
    []
    {\label{fig:topic6A}\includegraphics[width=.77\textwidth,height=3.4cm]{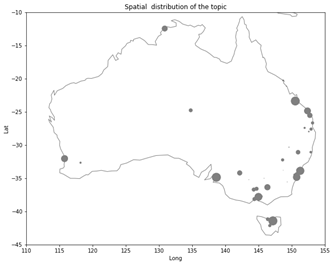}}
  \end{minipage}}
\usebox{\measurebox}\qquad
\begin{minipage}[b][\ht\measurebox][s]{.33\textwidth}
\centering
\subfloat
  []
  {\label{fig:topic6B}\includegraphics[width=.77\textwidth,height=1.5cm]{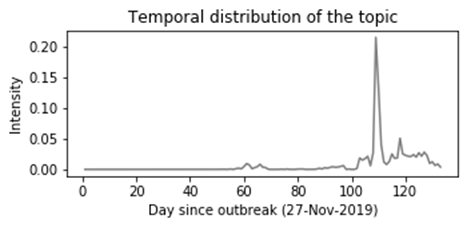}}
\vspace*{-1pt}
\subfloat
  []
  {\label{fig:topic6C}\includegraphics[width=.77\textwidth,height=1.5cm]{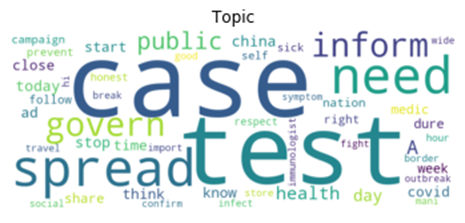}}
\end{minipage}
\caption{Topic 6. Summary: This topic is similar to Topic 5, but it is widespread across the nation, while Topic 4 is more focused on capital territory.}
\label{fig:topic6}
\end{figure*}

\begin{figure*}[ht!]
\centering
\sbox{\measurebox}{%
  \begin{minipage}[b]{.33\textwidth}
  \subfloat
    []
    {\label{fig:topic7A}\includegraphics[width=.77\textwidth,height=3.4cm]{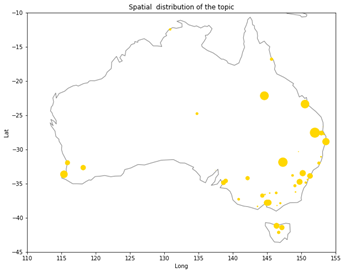}}
  \end{minipage}}
\usebox{\measurebox}\qquad
\begin{minipage}[b][\ht\measurebox][s]{.33\textwidth}
\centering
\subfloat
  []
  {\label{fig:topic7B}\includegraphics[width=.77\textwidth,height=1.5cm]{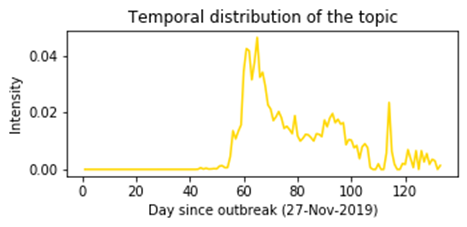}}
\vspace*{-1pt}
\subfloat
  []
  {\label{fig:topic7C}\includegraphics[width=.77\textwidth,height=1.5cm]{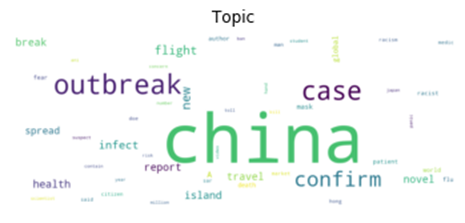}}
\end{minipage}
\caption{Topic 7. Summary: The Covid-19 outbreak in China and the earlier flight disruptions. This is the earliest pattern that emerged that gets to increase in a wider talk from 55$^{th}$ day (month of January) all over the nation. It is interesting to see how the topic suddenly dropped its peak when Covid-19 starts to hit Australia from 100$^{th}$ day.}
\label{fig:topic7}
\end{figure*}

\begin{figure*}[ht!]
\centering
\sbox{\measurebox}{%
  \begin{minipage}[b]{.33\textwidth}
  \subfloat
    []
    {\label{fig:topic8A}\includegraphics[width=.77\textwidth,height=3.4cm]{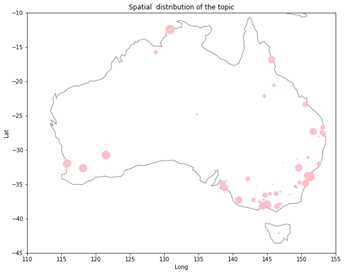}}
  \end{minipage}}
\usebox{\measurebox}\qquad
\begin{minipage}[b][\ht\measurebox][s]{.33\textwidth}
\centering
\subfloat
  []
  {\label{fig:topic8B}\includegraphics[width=.77\textwidth,height=1.5cm]{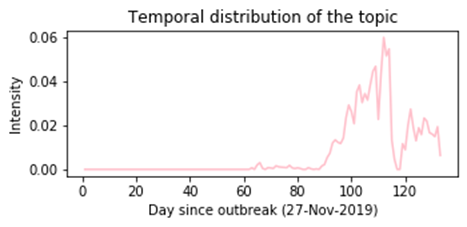}}
\vspace*{-1pt}
\subfloat
  []
  {\label{fig:topic8C}\includegraphics[width=.77\textwidth,height=1.5cm]{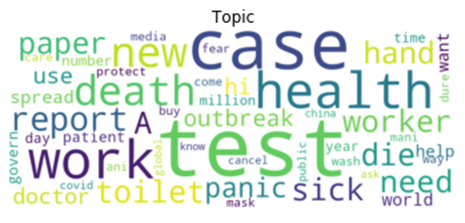}}
\end{minipage}
\caption{Topic 8. Summary: The start of panic buying toiler paper and Covid-19 case reports. The panic buying of toilet paper was seen everywhere in the nation and lasts for a while. Both the spatial and temporal patterns reflect it. \\}
\label{fig:topic8}
\end{figure*}

\begin{figure*}[ht!]
\centering
\sbox{\measurebox}{%
  \begin{minipage}[b]{.33\textwidth}
  \subfloat
    []
    {\label{fig:topic9A}\includegraphics[width=.77\textwidth,height=3.4cm]{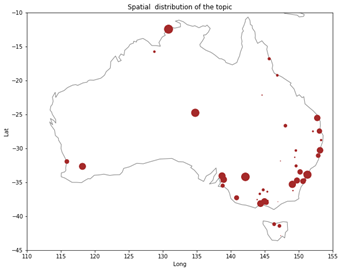}}
  \end{minipage}}
\usebox{\measurebox}\qquad
\begin{minipage}[b][\ht\measurebox][s]{.33\textwidth}
\centering
\subfloat
  []
  {\label{fig:topic9B}\includegraphics[width=.77\textwidth,height=1.5cm]{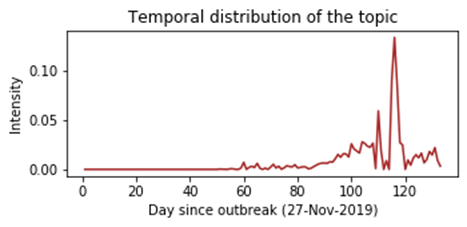}}
\vspace*{-1pt}
\subfloat
  []
  {\label{fig:topic9C}\includegraphics[width=.77\textwidth,height=1.5cm]{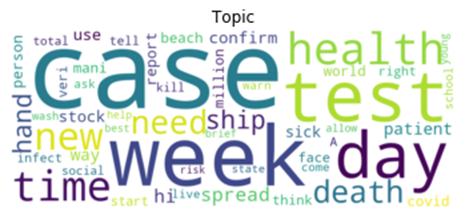}}
\end{minipage}
\caption{Topic 9. Summary: A general pattern on daily/weekly Covid-19 cases and tests conducted. It was steadily increasing from day 85 and starts to drop after the 115$^{th}$ day. This topic is also widespread across the nation.}
\label{fig:topic9}
\end{figure*}

\begin{figure*}[ht!]
\centering
\sbox{\measurebox}{%
  \begin{minipage}[b]{.33\textwidth}
  \subfloat
    []
    {\label{fig:topic10A}\includegraphics[width=.77\textwidth,height=3.4cm]{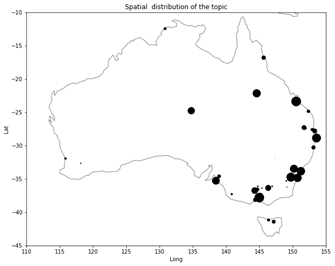}}
  \end{minipage}}
\usebox{\measurebox}\qquad
\begin{minipage}[b][\ht\measurebox][s]{.33\textwidth}
\centering
\subfloat
  []
  {\label{fig:topic10B}\includegraphics[width=.77\textwidth,height=1.5cm]{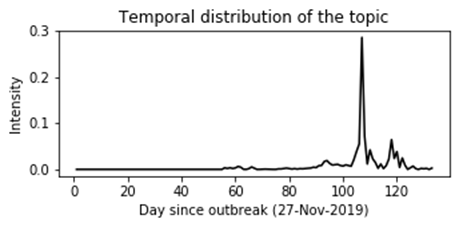}}
\vspace*{-1pt}
\subfloat
  []
  {\label{fig:topic10C}\includegraphics[width=.77\textwidth,height=1.5cm]{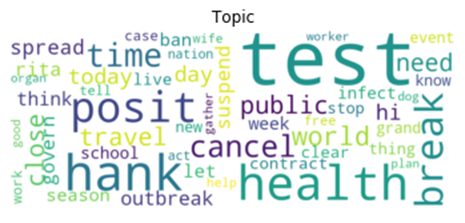}}
\end{minipage}
\caption{Topic 10. Summary: Hollywood actor Tom Hanks tested positive for Covid-19 in Gold coast, Australia. It is a short pattern that lasts only for a few days and a small peak again when he was cured. This topic is widely spread across all major cities in Australia.}
\label{fig:topic10}
\end{figure*}

Based on Fig.~\ref{patterns_ntf}a, Topics 3, 6, 7, and 10 can be identified as highly deviated ones or very popular ones for a short period of time. Similarly, Fig.~\ref{patterns_ntf}c shows some interesting spatial patterns like the green color (Topic 2) that is referring to a port on the western coast of Australia. So it could be linking to cruises that were one of the main Covid-19 clusters in Australia. While the main Australian cities like Sydney and Melbourne are showing many topics, a few topics are popular in other regional areas. Therefore, it is worth relating all these ten topics concerning spatial and temporal patterns and it is shown in Figs.~\ref{fig:topic1} to \ref{fig:topic10}. Figs.~\ref{fig:topic1A} to \ref{fig:topic10A} shows the spatial patterns of 10 topics with each pattern showing a distinctive patterns within Australia. By relating the topics (Figs.~\ref{fig:topic1C} to \ref{fig:topic10C}) with the spatial (\ref{fig:topic1A} to \ref{fig:topic10A}) and the temporal (\ref{fig:topic1B} to \ref{fig:topic10B}) patterns, more interesting insights can be identified. 

The main advantage of NTF is its ability to represent the associations among the modes. While existing techniques can elicit spatial or temporal patterns individually, they cannot preserve the association. Whereas in NTF, it is possible to associate the patterns. For example, in Topic 3, the sudden peak in the temporal pattern as shown in Fig.~\ref{patterns_ntf}a can be directly associated with the locations marked in Figs.~\ref{patterns_ntf}b and \ref{patterns_ntf}c.

%\vspace{-1em}

%\vspace*{20pt}

\section{Conclusion}
Understanding the spatio-temporal topic dynamics from social media can be helpful to make some decisions for governments or organizations. Especially with the Covid-19 crisis, a better understanding of people's behavior can help to take action by quickly grasping the people's expectations. Those actions can be very important like controlling panic buying, providing more medical and healthcare facilities, or focused attention on certain locations. The social media data representation and NTF based spatio-temporal topic dynamics extraction as proposed in this paper can bring a more sophisticated understanding of people concerning their space and time. This elicitation of patterns reduces much workload for users and highlights the associations automatically that are otherwise very difficult to identify by human observations. 
\bibliographystyle{IEEEtran}
\bibliography{main}
\end{document}